\documentclass[intlimits,twoside,a4paper]{article}

\usepackage[cp1251]{inputenc}

\usepackage{amsmath,amssymb}
\usepackage{graphicx}

\usepackage[eqsecnum]{cmpj3}

\usepackage{bm}

\issue{2019}{22}{2}{23703}
\doinumber{10.5488/CMP.22.23703}

\title[New exact results for two quasiparticles]%
{New exact analytical results for two quasiparticle excitation in the fractional quantum Hall effect}
\author{Z. Bentalha}
\address{
University of Tlemcen, Theoretical Physics Laboratory, 13000 Tlemcen, Algeria
}

\date{Received	March 20, 2019, in final form April 7, 2019}

\begin{document}

\maketitle

\begin{abstract}
In this work, two quasiparticle excitation energies per particle are calculated analytically for systems with up to $N = 7$ electrons in both Laughlin and composite fermions (CF) theories by considering the full jellium potential which consists of three parts, the electron-electron, electron-background, and background-background Coulomb interactions. The exact results we have obtained confirm the fact that the CF-wavefunction for two quasiparticles has lower energy than Laughlin wavefunction though the found difference between Laughlin and (CF) two quasiparticle energies decreases as the system size increases. 
\pacs 73.43.-f, 73.43.Cd
\end{abstract}

\section{Introduction}
\vspace{-2mm}
Many experiments have reported results that support the concept of fractionally charged quasiparticles in an electron gas under fractional quantum Hall effect conditions (see reference~\cite{R1} and references therein). These quasiparticles can be anyons, an exotic type of particle that is neither a fermion nor a boson~\cite{R2}. Composite fermions of Jain~\cite{R3} are also examples of such quasiparticles that are used in describing fractional quantum Hall effect (FQHE) ground states. Either Jain composite fermion wavefunction or Laughlin wavefunction~\cite{R4,R5} satisfactorilly describes FQHE ground states at the filling $\nu=1/m$, $m$ odd, to the point of being identical to each other. However, the wavefunctions for the excitations are different, which provides an opportunity to carry out a comparison between Jain and Laughlin wavefunctions even at the filling factor $\nu=1/3$. Thereafter in this paper, when speaking about quasiparticles (quasiholes), we will just mean quasiparticle (quasihole) excitations. In a previous study aimed at explaining the nature of quasiparticle excitations in the fractional quantum Hall effect~\cite{R6}, the authors of this investigation focused on the case $\nu=1/3$, and used only electron-electron Coulomb interaction. Their results for the two quasiparticle energy obtained by numerical Monte Carlo simulations revealed peculiar features, namely the wavefunction of Laughlin for two quasiparticles has much higher energy than Jain wavefunction, and in addition to that, the observed discrepancy increases as the system of electrons grows. These findings are also confirmed in reference~\cite{R7} via analytical calculations. In this paper, we again undertake the work of reference~\cite{R6} but by considering the full jellium potential which consists of three parts, the electron-electron, electron-background and background-background Coulomb interactions, then analytically calculate besides the energy of the two quasiholes, the energy of the two quasiparticles for both Laughlin and Jain theories at the filling $\nu=1/3$ for systems with up to $N=7$ electrons. The main concern of this work is to investigate, using only analytical calculations and focusing on the case $\nu=1/3$, the differences between Laughlin and Jain theories in describing the two quasiparticle excitation, to determine whether there is a large discrepancy between the two approaches, and whether this discrepancy increases as the system of electrons grows as it is reported in reference~\cite{R6}. In a broader perspective, the topic of quasiparticle excitations can be used in anyonic or non-abelian exchange statistics \cite{R8,R9,R10,R11} that are seen as promising areas of research especially in topological quantum computation (TQC) \cite{R12,R13}. 

The paper is organized as follows. In section \ref{sec2}, a theoretical setting is presented. In section\ref{sec3}, details about quasiparticle and quasihole excitation energies are presented clarifying their inherent definitions. In section~\ref{sec4}, we present the method of calculation of the energies for the single quasiparticle and quasihole excitations. In section\ref{sec5}, the two quasiparticle and quasihole energies are derived. Some concluding remarks are given in section~\ref{sec6}.     
\section{The model}\label{sec2}
Let us consider $N ( > 2)$ electrons of charge ($-e$) embedded in a uniform neutralizing background disk of positive charge $N e$ and area $S=\piup R^{2}$, where $R$ represents the radius of the disk. This 2D electronic system is subjected to a strong perpendicular uniform magnetic field  ${\bf{B}}$ in the $z$ direction, and the underlying physics is governed by the full jellium interaction potential $V$,
\begin{equation}
V=V_\text{ee}+V_\text{eb}+V_\text{bb}\,,
\label{E1}
\end{equation}
where $V_\text{ee}, V_\text{eb}$ and $V_\text{bb}$ are the electron-electron, electron-background and background-background interaction potentials, respectively. Their corresponding expressions are as follows:
\begin{eqnarray}
V_\text{ee}&=&\sum_{i<j}^{N}\frac{e^{2}}{\mid {\bf{r}}_{i}- {\bf{r}}_{j} \mid}\,,
\label{E2}
\end{eqnarray}
\begin{eqnarray}
V_\text{eb}&=&-\rho\sum_{i=1}^{N}\int_{S}\rd^{2}r\frac{e^{2}}{\mid {\bf{r_{i}}}- {\bf{r}} \mid}\,,
\label{E3}
\end{eqnarray}
\begin{eqnarray}
V_\text{bb}&=&\frac{\rho^{2}}{2}\int_{S} \rd^{2}r \int_{S} \rd^{2}r^{\prime}\frac{e^{2}}{\mid {\bf{r}}- {\bf{r^{\prime}}} \mid}\,,
\label{E4}
\end{eqnarray}     
where ${\bf{r}}_{i}$ (or ${\bf{r}}_{j}$) denotes the electron vector position while ${\bf{r}}$ and ${\bf{r^{\prime}}}$ are background coordinates. $S(B)$ is the area of the disk and $\rho(B)$ is the density of the system (the number of electrons per unit area) that can also be defined as
\begin{equation}
\rho=\frac{\nu}{2\piup l^{2}}\,,
\label{E5}
\end{equation}    
where $l(B)=\sqrt{{\hbar\,c}/{e B}}$ is the magnetic length, $c$ is the speed of light, $B$ is the magnetic field strength and $\nu$ is the filling factor. The background-background interaction potential can be calculated classically without using the wave function of the electron system. Its value is determined analytically \cite{R14} and is given by 
\begin{equation}
V_\text{bb}=\frac{8}{3\piup}\frac{e^{2}(\rho S)^{2}}{R}.
\label{E6}
\end{equation}
Henceforth our concern will be only directed to $V_\text{ee}$ and $V_\text{eb}$. For a given wavefunction $\psi_{\chi}({\bf{r}}_{1},{\bf{r}}_{2},\dots,{\bf{r}}_{N})$, where $\chi$ can be either L or CF to designate Laughlin or composite fermion (CF) description, the electron-electron and  electron-background interaction energies are written as \cite{R14},
\begin{eqnarray}
\left\langle V_\text{ee}\right\rangle_{\chi} = \frac{N(N-1)}{2 \, Z}
\int \rd^{2}r_{1}\dots \rd^{2}r_{N} \frac{e^{2}}{\mid {\bf{r}}_{1}- {\bf{r}}_{2} \mid} \mid \psi_{\chi}({\bf{r}}_{1},\dots,{\bf{r}}_{N})\mid^{2},
\label{E7}
\end{eqnarray}
\begin{eqnarray}
\langle V_\text{eb}\rangle_{\chi} = -\frac{\rho N}{Z}
\int \rd^{2}r_{1}\dots\rd^{2}r_{N}\mid \psi_{\chi}({\bf{r}}_{1},\dots,{\bf{r}}_{N})\mid^{2} \int_{S}\rd^{2}r \frac{e^{2}}{\mid {\bf{r}}_{1}- {\bf{r}} \mid}\,,
\label{E8}
\end{eqnarray}
where $Z$ is nothing else than the norm, that is $Z=\int \rd^{2}r_{1}\dots \rd^{2}r_{N}\mid \psi_{\chi}({\bf{r}}_{1},\dots,{\bf{r}}_{N})\mid^{2}$. Taking into account the fact that \cite{R15}
\begin{equation}
\int_{S}\rd^{2}r \frac{e^{2}}{\mid {\bf{r}}_{1}- {\bf{r}} \mid}=2\frac{e^{2} S}{R} \int_{0}^{\infty} \frac{\rd q}{q}J_{1}(q)J_{0}\Big(\frac{q}{R}r_{1}\Big)
\label{E9}
\end{equation}
the expression of $\langle V_\text{eb}\rangle_{\chi}$ can also be written as follows:
\begin{eqnarray}
\langle V_\text{eb}\rangle_{\chi}=-\frac{2 e^{2} N\, (\rho S)}{R \, Z} \int_{0}^\infty \rd q \frac{J_{1}(q)}{q} 
\int \rd^{2}r_{1}\dots \rd^{2}r_{N}\mid \psi_{\chi}({\bf{r}}_{1},\dots,{\bf{r}}_{N})\mid^{2} J_{0}\Big(\frac{q r_{1}}{R}\Big)\,,
\label{E10}
\end{eqnarray}
where $J_{n}(x)$ are $n$-th order Bessel functions. In a complex notation, equations~(\ref{E7}) and (\ref{E10}) transform into,
\begin{eqnarray}
\left\langle V_\text{ee}\right\rangle_{\chi} = \frac{N(N-1)}{2 \, Z}
\int \rd^{2}z_{1}\dots \rd^{2}z_{N} \frac{e^{2}}{\mid z_{1}- z_{2}\mid} \mid \psi_{\chi}(z_{1},\dots,z_{N})\mid^{2},
\label{E11}
\end{eqnarray}
\begin{eqnarray}
\langle V_\text{eb}\rangle_{\chi} =-\frac{2 e^{2} N\,(\rho S)}{R \, Z} \int_{0}^{\infty} \rd q \frac{J_{1}(q)}{q} 
\int \rd^{2}z_{1}\dots \rd^{2}z_{N}\mid \psi_{\chi}(z_{1},\dots,z_{N})\mid^{2} J_{0}\bigg(\frac{q\mid z_{1}\mid}{R}\bigg).
\label{E12}
\end{eqnarray}
Moreover, in order to investigate the differences between two descriptions for the quasiparticle, we analytically calculate the expectation values of $\left\langle V_\text{ee}\right\rangle_{\chi}$ and $\langle V_\text{eb}\rangle_{\chi}$ for each $\chi$-description. The energy $V_\text{bb}$ is of the same value in both theories because it has no dependence on the wave function. The excited state energy per particle is defined as $\bar{\varepsilon}_{\chi}=\bar{\varepsilon}_{\text{ee}\chi}+\bar{\varepsilon}_{\text{eb}\chi}+\bar{\varepsilon}_\text{bb}$ where $\bar{\varepsilon}_{\chi}= \langle V\rangle_{\chi}/N$, $\bar{\varepsilon}_{\text{ee}\chi}= \langle V_\text{ee}\rangle_{\chi}/N$, $\bar{\varepsilon}_{\text{eb}\chi}= \langle V_\text{eb}\rangle_{\chi}/N$, and $\bar{\varepsilon}_\text{bb}= V_\text{bb}/N$ are, respectively, the total, (e-e), (e-b), and (b-b) energies per particle in the $\chi$-description. The wave function of the quasihole is the same for the two theories, hence its energy has no impact on comparing the two theories, but its value is crucial to compute the energy gap of the first or second excited state.

\section{Definition of the quasiparticle and quasihole}\label{sec3}
Let us define the energy of the quasiparticle. To this end, we follow the lines of reference~\cite{R16} wherein two definitions are proposed for the quasiparticle energy, the gross and the proper energies which evidently lead to identical results for the energy gap. In this work, we adopt the proper energy as a definition for the quasiparticle. We start with a ground state of $N$ electrons at the filling $\nu=1/3$ immersed in a strong magnetic field $B_{0}$, then we keep $N$ constant and reduce the magnetic field by a factor of $(mN-1/mN)$, with $m=1/\nu$. As one has $l(B)=\sqrt{{\hbar\,c}/{e B}}$, the magnetic length will be increased by the factor $\sqrt{{mN}/{(mN-1)}}$, that is $l(B)=\sqrt{{mN}/{(mN-1)}}l_{0}$ with $l_{0}=\sqrt{{\hbar\,c}/{e B_{0}}}$, thus the outer radius of the occupied electron disk is the same as the original ground state leading to a reduction in the uniform electron density elsewhere in the disk in proportion to the reduction of $B$ according to $\rho(B)={1}/{2 \piup l^{2}(B)}$. The quasiparticle energy is defined as the difference between the energies of excited and ground states of systems with the same number of particles $N$ as well as the same physical area $S_{0}$, and slightly different values of the magnetic field~\cite{R16}. Thus, we refer to $\varepsilon^{-}_{\chi}$ as the quasiparticle energy per particle in the $\chi$-description which reflects the change in potential energy upon removing one quantum of the magnetic flux from the system, i.e.,
\begin{equation}
\varepsilon^{-}_{\chi}=\bar{\varepsilon}^\text{qp}_{\chi}(S_{0})-\varepsilon(S_{0})\,,
\label{E13}
\end{equation} 
where $\varepsilon$ is the total ground state energy per particle at $\nu=1/3$ which is calculated analytically for various values of $N$ in reference~\cite{R17}, its corresponding wave function is~\cite{R5}, 
\begin{equation}
\psi_\text{GS}= \re^{-\sum_{k}\frac{\mid z_{k}\mid^{2}}{4}}\:\prod\limits_{i<j}^{N}(z_{i}-z_{j})^{3}, 
\label{E14}
\end{equation}
where distances are measured in units of the magnetic length $l_{0}$, unless otherwise noted, hence $S_{0}=\piup(2mN)$. Similarly, for the quasihole energy, upon adding one quantum of magnetic flux to the system, the magnetic field is increased by a factor of $(mN+1)/mN$, the magnetic length is then reduced by the factor of $mN/(mN+1)$, and the outer radius is maintained identical to the original ground state radius, the quasihole energy is then given by
\begin{equation}
\varepsilon^{+}=\bar{\varepsilon}^\text{qh}(S_{0})-\varepsilon(S_{0})\,,
\label{E15}
\end{equation}
as underlined above, the quasihole energy is the same in both theories. 
\section{The single quasiparticle and quasihole}\label{sec4}
Let us start by giving the formulae of the quasiparticle wavefunctions in both theories. For Laughlin, the effect of piercing the quantum fluid at the origin with an infinitely thin solenoid and removing through it a flux quantum $\phi_{0}=hc/e$ adiabatically motivates the following wavefunction for the quasiparticle~\cite{R5},
\begin{equation}
\psi^\text{qp}_\text{1L}= \re^{-\sum_{k}\frac{\mid z_{k}\mid^{2}}{4l^{2}}}\:\prod\limits_{k}\bigg(2\,\frac{\partial}{\partial z_{k}}\bigg)\prod\limits_{i<j}^{N}(z_{i}-z_{j})^{3}. 
\label{E16}
\end{equation}
In CF theory, the single quasiparticle wavefunction is given by\,\cite{R6},
\begin{equation}
\psi^\text{qp}_\text{1CF}=\sum_{i=1}^{N}\frac{\sum_{k}^{\prime}(z_{i}-z_{k})^{-1}}{\prod\limits_{j}\!^{\prime}(z_{j}-z_{i})}\:\psi_\text{GS}(l)\,,
\label{E17}
\end{equation}
where the prime denotes the condition $j\neq i$ or $k\neq i$ and $\psi_\text{GS}(l)$ is the $\nu=1/3$ ground state wave function at magnetic length $l(B)$. The single quasiparticle energy is defined by~\cite{R16},
\begin{equation}
\varepsilon^{-}_{1\chi}=\bar{\varepsilon}^\text{qp}_{1\chi}(S_{0})-\varepsilon(S_{0})\,,
\label{E18}
\end{equation}
as noted in the definition of the quasiparticle energy~\cite{R16}, 
\begin{equation}
\bar{\varepsilon}^\text{qp}_{1\chi}(S_{0})=\left( \frac{S}{S_{0}} \right)^{1/2}\bar{\varepsilon}^\text{qp}_{1\chi}(S)\,,
\label{E19}
\end{equation}
with $S=\piup \, 2(Nm-1)$ as though $l$ is set to one and,
\begin{eqnarray}
\bar{\varepsilon}^\text{qp}_{1\chi}(S)& = & \frac{\langle V_\text{ee}\rangle_{1\chi}(S)+\langle V_\text{eb}\rangle_{1\chi}(S)+V_\text{1bb}(S)}{N} \nonumber \\
& = & \bar{\varepsilon}^\text{qp}_{1\text{ee}\chi} + (x_{1}^{-})^{\frac{1}{2}}\,\bar{\varepsilon}^\text{qp}_{1\text{eb}\chi} +(x_{1}^{-})^{\frac{3}{2}}\, \bar{\varepsilon}_\text{bb}\,, \label{E20} 
\end{eqnarray}
where various energies per particle are as follows,
\begin{eqnarray}
\bar{\varepsilon}^\text{qp}_{1\text{ee}\chi} = \frac{(N-1)}{2 \, Z_{1\text{qp}}}
\int \rd^{2}z_{1}\dots \rd^{2}z_{N} \frac{e^{2}}{\mid z_{1}- z_{2}\mid} \mid \psi^\text{qp}_{1\chi}(z_{1},\dots,z_{N})\mid^{2},
\label{E21}
\end{eqnarray}
\begin{eqnarray}
\bar{\varepsilon}^\text{qp}_{1\text{eb}\chi} =-\frac{2 e^{2}}{Z_{1\text{qp}}}\sqrt{\frac{N}{2m}} \int_{0}^{\infty} \rd q \frac{J_{1}(q)}{q} 
\int \rd^{2}z_{1}\dots \rd^{2}z_{N}\mid \psi^\text{qp}_{1\chi}(z_{1},\dots,z_{N})\mid^{2} J_{0}\bigg(\frac{q\mid z_{1}\mid}{R}\bigg),
\label{E22}
\end{eqnarray}
\begin{equation}
\bar{\varepsilon}_\text{bb}=\frac{8}{3\piup}e^{2}\sqrt{\frac{N}{2m}}\,,
\label{E23}
\end{equation}
the factor $x_{1}^{-}=1-{1}/{mN}$, and $R=\sqrt{2(mN-1)}$. We know that the factor $x_{1}^{-}$ tends to one for very large size systems (realistic systems), but as in most many-body analytical calculations, we are limited to small systems of several electrons. This situation can be improved by considering the following formula for the quasiparticle energy (to ovoid the energies less than or nearly ground state energies),
\begin{equation}
\bar{\varepsilon}^\text{qp}_{1\chi}(S_{0})= \left[\bar{\varepsilon}^\text{qp}_{1\text{ee}\chi} + (x_{1}^{-})^{1/2}\,\bar{\varepsilon}^\text{qp}_{1\text{eb}\chi} +\bar{\varepsilon}_\text{bb}\right](x_{1}^{-})^{1/2}.
\label{E24}
\end{equation}
The wavefunction of the quasihole at the origin is given by~\cite{R5},
\begin{equation}
\psi^\text{qh}_{1}= \re^{-\sum_{k}\frac{\mid z_{k}\mid^{2}}{4l^{2}}}\:\prod\limits_{k}( z_{k})\prod\limits_{i<j}^{N}(z_{i}-z_{j})^{3}. 
\label{E25}
\end{equation}
In this case, we adopt the following expression for the single quasihole energy,
\begin{equation}
\bar{\varepsilon}^\text{qh}_{1\chi}(S_{0})= \left[\bar{\varepsilon}^\text{qh}_{1\text{ee}} + (x_{1}^{+})^{1/2}\,\bar{\varepsilon}^\text{qh}_{1\text{eb}} +(x_{1}^{+})^{3/2}\,\bar{\varepsilon}_\text{bb}\right](x_{1}^{+})^{1/2},
\label{E26}
\end{equation}
with $x_{1}^{+}=1+{1}/{mN}$,
\begin{eqnarray}
\bar{\varepsilon}^\text{qh}_{1\text{ee}} = \frac{(N-1)}{2 \, Z_\text{1qh}}
\int \rd^{2}z_{1}\dots \rd^{2}z_{N} \frac{e^{2}}{\mid z_{1}- z_{2}\mid} \mid \psi^\text{qh}_{1}(z_{1},\dots,z_{N})\mid^{2},
\label{E27}
\end{eqnarray}
\begin{eqnarray}
\bar{\varepsilon}^\text{qh}_\text{1eb} =-\frac{2 e^{2}}{Z_\text{1qh}}\sqrt{\frac{N}{2m}} \int_{0}^{\infty} \rd q \frac{J_{1}(q)}{q} 
\int \rd^{2}z_{1}\dots \rd^{2}z_{N}\mid \psi^\text{qh}_{1}(z_{1},\dots,z_{N})\mid^{2} J_{0}\bigg(\frac{q\mid z_{1}\mid}{R}\bigg),
\label{E28}
\end{eqnarray}
and $R=\sqrt{2(mN+1)}$. The results are given in the subsequent paragraph.
\subsection*{The results for the single quasiparticle and quasihole}
We draw up three tables corresponding to various energies for the quasiparticle and quasihole, then a comparison is made between the wavefunctions of Laughlin and Jain.  
\begin{table}[!b]
\caption{\label{Table1} The single quasiparticle excitation energies (in units of $e^{2}/l_{0}$) are given for the Laughlin wavefunction at the filling $\nu={1}/{3}$ for systems with up to $N= 7$ electrons. In the sixth column are given the energies of the ground state at the filling $\nu = 1/3$ reported in references~\cite{R17, R18}.}
\vspace{2ex}
\begin{center}
\begin{tabular}{lllllll}
  \hline 
  \hline
  \\                             
  $N$ & $\bar{\varepsilon}_\text{bb}$ & $ \bar{\varepsilon}^\text{qp}_\text{1ebL}$ & $ \bar{\varepsilon}^\text{qp}_\text{1eeL}$ & $ \bar{\varepsilon}^\text{qp}_\text{1L}(S_{0})$ &  $\varepsilon (S_{0})$     & $ \varepsilon^{-}_\text{1L}$ \\ 
  \\
 \hline
   4 & 0.693064 & -- 1.437430  & 0.375346      & -- 0.294719  & -- 0.388855     & 0.094136 \\
   5 & 0.774869 & -- 1.595668  & 0.449748      & -- 0.306198  & -- 0.390255     & 0.084057 \\ 
   6 & 0.848826 & -- 1.739305  & 0.518946      & -- 0.313441  & -- 0.391517     & 0.078076 \\
   7 & 0.916837 & -- 1.871819  & 0.582871      & -- 0.31912   & -- 0.392624     & 0.073504  \\ 
   \hline
\end{tabular}
\end{center}
\end{table}
\begin{table}[!h]
\caption{\label{Table2} The single quasiparticle excitation energies (in units of $e^{2}/l_{0}$) are given for the (CF)-wavefunction at the filling $\nu={1}/{3}$ for systems with up to $N= 7$ electrons.}
\vspace{2ex}
\begin{center}
\begin{tabular}{lllllll}
  \hline 
  \hline
  \\                             
  $N$ & $\bar{\varepsilon}_\text{bb}$ & $ \bar{\varepsilon}^\text{qp}_\text{1ebCF}$ & $ \bar{\varepsilon}^\text{qp}_\text{1eeCF}$ & $ \bar{\varepsilon}^\text{qp}_\text{1CF}(S_{0})$ &  $\varepsilon (S_{0})$     & $ \varepsilon^{-}_\text{1CF}$ \\ 
  \\
  \hline
   4 & 0.693064 & -- 1.437074  & 0.37543     & -- 0.294313  & -- 0.388855   & 0.094542 \\
   5 & 0.774869 & -- 1.594935  & 0.449808(5) & -- 0.305455  & -- 0.390255   & 0.0848 \\ 
   6 & 0.848826 & -- 1.738382  & 0.5187      & -- 0.312809  & -- 0.391517   & 0.078708 \\
   7 & 0.916837 & -- 1.870889  & 0.582317    & -- 0.318775  & -- 0.392624   & 0.073849  \\ 
   \hline
\end{tabular}
\end{center}
\end{table}
\begin{table}[!h]
\caption{\label{Table3} The single quasihole excitation energies (in units of $e^{2}/l_{0}$) are given for systems with up to $N= 7$ electrons at the filling $\nu={1}/{3}$.}
\vspace{2ex}
\begin{center}
\begin{tabular}{lllllll}
  \hline 
  \hline
  \\                             
  $N$ & $\bar{\varepsilon}_\text{bb}$ & $ \bar{\varepsilon}^\text{qh}_\text{1eb}$ & $ \bar{\varepsilon}^\text{qh}_{1\text{ee}}$ & $ \bar{\varepsilon}^\text{qh}_{1}(S_{0})$ &  $\varepsilon (S_{0})$     & $ \varepsilon^{+}_{1}$ \\ 
  \\
  \hline
   4 & 0.693064  & -- 1.388968   & 0.296363 & -- 0.355742 & -- 0.388855   & 0.033113 \\
   5 & 0.774869 & -- 1.553198  & 0.376761(5) & -- 0.361385 & -- 0.390255    & 0.02887 \\ 
   6 & 0.848826 & -- 1.701311  & 0.450376(5) & -- 0.364678  & -- 0.391517  & 0.026839 \\ 
   7 & 0.916837 & -- 1.837358  & 0.518071 & -- 0.367234 & -- 0.392624   &  0.02539\\ 
   \hline
\end{tabular}
\end{center}
\end{table}
We noticed that the wavefunction of Laughlin for the single quasiparticle has a lower energy than the (CF)-wavefunction, but this is only a feature of small size systems for two reasons. First, if we look carefully at the fourth column in tables~\ref{Table1} and \ref{Table2}, it can be realized that, for $N=6\,, 7$ electrons, the (CF)-wavefunction for the single quasiparticle has lower (e-e) interaction energy than Laughlin wavefunction, and for $N$ large, only the (e-e) interaction energy is crucial for the asymptotic limit value while the electron-background (e-b) and (b-b) interaction energies cancel because they are divergent terms with respect to $N$ together with the divergent part of the (e-e) interaction energy. Second, if we look at the difference $(\varepsilon^{-}_\text{1CF}-\varepsilon^{-}_\text{1L})$, it can be easily noticed that $(\varepsilon^{-}_\text{1CF}-\varepsilon^{-}_\text{1L})(N=5)\approx 0.0008$, $(\varepsilon^{-}_\text{1CF}-\varepsilon^{-}_\text{1L})(N=6)\approx 0.0007$, and $(\varepsilon^{-}_\text{1CF}-\varepsilon^{-}_\text{1L})(N=7)\approx 0.0003$, that is the energy difference is decreasing with increasing $N$. As concerns the energy gap, from tables~\ref{Table1}, \ref{Table2}, and \ref{Table3}, it can be verified that, for $N=7$ electrons, the energy gap for both Laughlin and Jain (CF) wavefunctions is nearly $0.099$.

\section{Two quasiparticles and quasiholes}\label{sec5}
\begin{figure}[!b]
	\begin{center}
		\includegraphics[height=6.0cm,width=10.0cm]{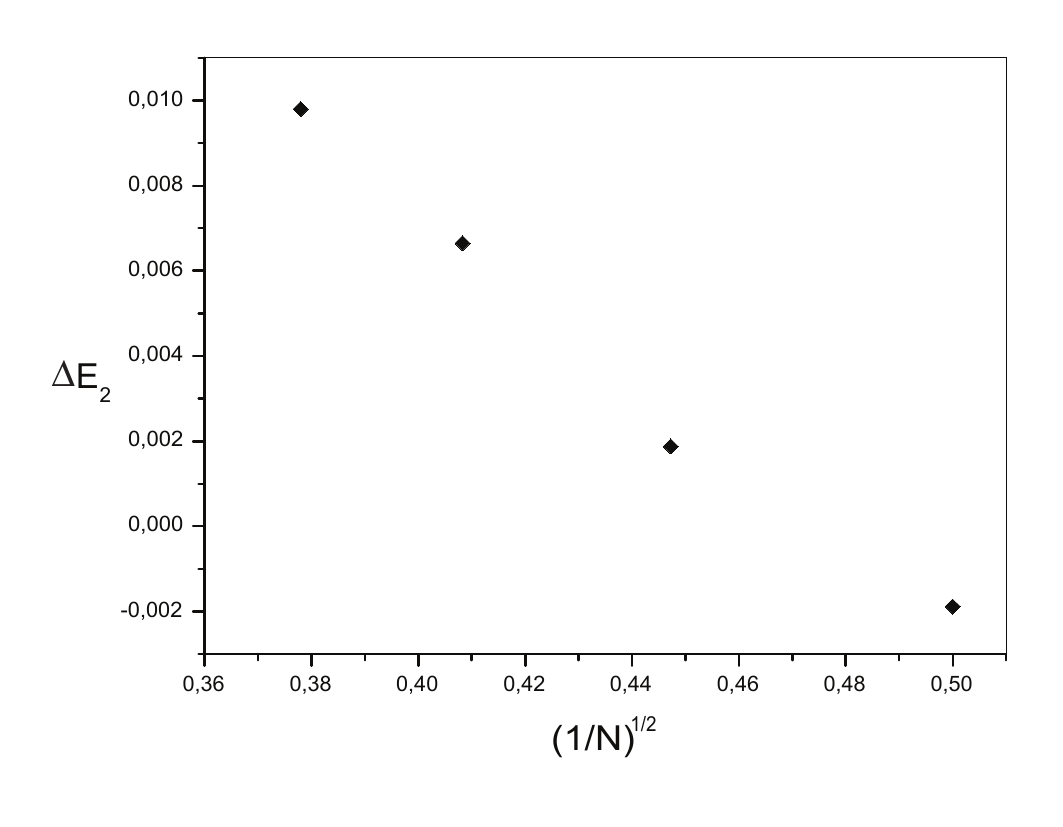}
	\end{center}
	\caption{\label{Fig.1} The difference $\Delta E_{2}$ between the energies of both Laughlin and Jain wavefunctions for two quasiparticles is plotted as a function of ${1}/{\sqrt{N}}$ (in units of ${e^{2}}/{l_{0}}$).}
\end{figure}

In the case of two quasiparticles, we also have two different wavefunctions. For Laughlin, the effect of removing the two quantum flux at the origin enhances the following wavefunction~\cite{R6}, 
\begin{equation}
\psi_\text{2L}^\text{qp}= \re^{-\sum_{k}\frac{\mid z_{k}\mid^{2}}{4l^{2}}}\:\prod\limits_{k}\bigg(2\, \frac{\partial}{\partial z_{k}}\bigg)\bigg(2\, \frac{\partial}{\partial z_{k}}\bigg)\prod\limits_{i<j}^{N}(z_{i}-z_{j})^{3} 
\label{E29}
\end{equation}
as a generalization of the single quasiparticle wavefunction. However, the (CF)-theory gives the following wavefunction for two quasiparticles at the origin~\cite{R6},
\begin{equation}
\psi_\text{2CF}^\text{qp}=\re^{-\sum_{k}\frac{\mid z_{k}\mid^{2}}{4l^{2}}} \, \mathcal{P} \prod\limits_{i<j}^{N}(z_{i}-z_{j})^{3} \, \begin{array}{|ccc|}
z_{1}^{*} & z_{2}^{*} & \cdots \\
z_{1}^{*}z_{1}^{*} & z_{2}^{*}z_{2}^{*} & \cdots \\
1 & 1 & \cdots \\
z_{1} & z_{2} & \cdots \\
\vdots & \vdots & \cdots \\
z_{1}^{N-3} & z_{2}^{N-3} & \cdots \\ 
\end{array}\,\,.
\label{E30}
\end{equation} 
It can be verified that the (CF)-wavefunction for two quasiparticles can be expressed as~\cite{R7},
\begin{equation}
\psi_\text{2CF}^\text{qp}=\sum_{i<j}\frac{\big(2\, \frac{\partial}{\partial z_{i}}\big)\big(2\, \frac{\partial}{\partial z_{j}}\big)}{\prod\limits_{k}\!^{\prime}(z_{k}-z_{i})(z_{k}-z_{j})}\:\psi_\text{GS}\,,
\label{E31}
\end{equation}
where the prime denotes the condition $k\neq i$ and $k\neq j$. Similarly, the wavefunction of the two quasiholes at the origin is given by,
\begin{equation}
\psi^\text{qh}_{2}= \re^{-\sum_{k}\frac{\mid z_{k}\mid^{2}}{4l^{2}}}\:\prod\limits_{k}( z_{k})( z_{k})\prod\limits_{i<j}^{N}(z_{i}-z_{j})^{3}, 
\label{E32}
\end{equation}
as a generalisation of Laughlin formula for the single quasihole~\cite{R5}.
Using equations (\ref{E11}) and (\ref{E12}) to compute the expectation values of various interactions, we obtain the energies of the second excited state (with two quasiparticles / two quasiholes at the origin) for both Laughlin and CF theories. The results are given in the tables below. One can observe from tables \ref{Table4} and \ref{Table5} that the (CF)-wavefunction for two quasiparticles has lower energy than Laughlin wavefunction. We expect that this feature remains even for large $N$ because in this case, also the (e-e) interaction in Jain (CF)-description has lower energy than that in Laughlin theory. The variation of the difference $(\varepsilon^{-}_\text{2L}-\varepsilon^{-}_\text{2CF})$ with $({1}{\sqrt{N}})$ is plotted in figure~\ref{Fig.1}. Similarly, from tables~\ref{Table4}, \ref{Table5}, and \ref{Table6}, it can be derived that, for $N=7$ electrons, the second excited state energy gap is nearly $0.19$ for the (CF)-wavefunction and $0.2$ for Laughlin wavefunction.

\begin{table}[!b]
\caption{\label{Table4} The two quasiparticle excitation energies (in units of $e^{2}/l_{0}$) are given for the wavefunction of Laughlin at the filling $\nu={1}/{3}$ for systems with up to $N= 7$ electrons.}
\vspace{2ex}
\begin{center}
\begin{tabular}{lllllll}
  \hline 
  \hline
  \\                             
  $N$ & $\bar{\varepsilon}_\text{bb}$ & $ \bar{\varepsilon}^\text{qp}_\text{2ebL}$ & $ \bar{\varepsilon}^\text{qp}_\text{2eeL}$ & $ \bar{\varepsilon}^\text{qp}_\text{2L}(S_{0})$ &  $\varepsilon (S_{0})$     & $ \varepsilon^{-}_\text{2L}$ \\ 
  \\
 \hline
   4 & 0.693064 & -- 1.469286  & 0.446495      & -- 0.184135   & -- 0.388855     & 0.20472 \\
   5 & 0.774869 & -- 1.623265  & 0.507621      & -- 0.212896   & -- 0.390255     & 0.177359\\ 
   6 & 0.848826 & -- 1.764095  & 0.57036       & -- 0.230063   & -- 0.391517     & 0.161454 \\
   7 & 0.916837 & -- 1.894250  & 0.630838      & -- 0.241713   & -- 0.392624     & 0.150911  \\ 
   \hline
\end{tabular}
\end{center}
\end{table}
\begin{table}[!b]
\caption{\label{Table5} The two quasiparticle excitation energies (in units of $e^{2}/l_{0}$) are given for the (CF)-wavefunction at the filling $\nu={1}/{3}$ for systems with up to $N= 7$ electrons.}
\vspace{2ex}
\begin{center}
\begin{tabular}{lllllll}
  \hline 
  \hline
  \\                             
  $N$ & $\bar{\varepsilon}_\text{bb}$ & $ \bar{\varepsilon}^\text{qp}_\text{2ebCF}$ & $ \bar{\varepsilon}^\text{qp}_\text{2eeCF}$ & $ \bar{\varepsilon}^\text{qp}_\text{2CF}(S_{0})$ &  $\varepsilon (S_{0})$     & $ \varepsilon^{-}_\text{2CF}$ \\ 
  \\
  \hline
   4 & 0.693064 & -- 1.439602  & 0.421462     & -- 0.18225   & -- 0.388855   & 0.206605 \\
   5 & 0.774869 & -- 1.602302  & 0.486087     & -- 0.214776  & -- 0.390255   & 0.175479 \\ 
   6 & 0.848826 & -- 1.748566  & 0.548666     & -- 0.236713  & -- 0.391517   & 0.154804 \\
   7 & 0.916837 & -- 1.882515  & 0.609375     & -- 0.251511  & -- 0.392624   & 0.141113  \\ 
   \hline
\end{tabular}
\end{center}
\end{table}
\begin{table}[!t]
\caption{\label{Table6} The two quasihole excitation energies (in units of $e^{2}/l_{0}$) are given for systems with up to $N= 7$ electrons at the filling $\nu={1}/{3}$.}
\vspace{2ex}
\begin{center}
\begin{tabular}{lllllll}
  \hline 
  \hline
  \\                             
  $N$ & $\bar{\varepsilon}_\text{bb}$ & $ \bar{\varepsilon}^\text{qh}_\text{2eb}$ & $ \bar{\varepsilon}^\text{qh}_\text{2ee}$ & $ \bar{\varepsilon}^\text{qh}_{2}(S_{0})$ &  $\varepsilon (S_{0})$     & $ \varepsilon^{+}_{2}$ \\ 
  \\
  \hline
   4 & 0.693064  & -- 1.370343   & 0.273240 & -- 0.311646 & -- 0.388855   & 0.077209 \\
   5 & 0.774869 & -- 1.536527  & 0.353251 & -- 0.324794 & -- 0.390255    & 0.065461 \\ 
   6 & 0.848826 & -- 1.686084  & 0.427601 & -- 0.332401  & -- 0.391517  & 0.059116 \\ 
   7 & 0.916837 & -- 1.823170  & 0.496439 & -- 0.337588 & -- 0.392624   &  0.055036\\ 
   \hline
\end{tabular}
\end{center}
\end{table}

\section{Concluding remarks}\label{sec6}
In this work we analytically calculated the quasiparticule energies per particle for both Laughlin and Jain (CF) theories for the most stable FQHE state which is the $\nu=1/3$ state. The results concerning the single quasiparticle energies obtained in each theory sustain the idea that Jain (CF)-wavefunction for the single quasiparticle has a lower energy than Laughlin wavefunction. Moreover, it has been observed in reference~\cite{R6} that $\psi^\text{qp}_\text{2L}$ has much higher energy than $\psi^\text{qp}_\text{2CF}$, and the noted discrepancy increases as the system of electrons grows. The results of this work confirm the feature that $\psi^\text{qp}_\text{2CF}$ has lower energy than~$\psi^\text{qp}_\text{2L}$, but do not sustain the idea that the energy difference increases as the system size grows, because it can be noticed that $[\Delta E_{2}(N=6)-\Delta E_{2}(N=5)]=0.00477$ and $[\Delta E_{2}(N=7)-\Delta E_{2}(N=6)]=0.003148$, that is the difference is decreasing as $N$ augment as shown in the figure above. In spite of that, it would be interesting to push the analytical calculations to $N > 7$ electrons in order to better clarify the result of comparison between Laughlin and Jain wavefunctions for the quasiparticle excitations.

\newpage
\ukrainianpart

\title{Нові точні аналітичні результати для збудження двох квазічастинок при дробовому квантовому ефекті Гола}
\author{З. Бентала}
\address{
	Університет Тлемсена, лабораторія теоретичної фізики, 13000 Тлемсен, Алжир	
}

\makeukrtitle

\begin{abstract}
	У даній роботі здійснено аналітичне обчислення  енергій  двохквазічастинкове збудження  для систем з  $N = 7$ електронів у відповідності до теорiї Лафлiна та теорiї 
	складних фермiонiв, враховуючи повний потенціал желе, який складається з трьох частин, кулонівських взаємодій електрон-електрон,  електрон-фон та фон-фон. Точні результати, отримані в даній роботі, підтверджують, що хвильова функція складних ферміонів для двох квазічастинок має меншу енергію, ніж хвильова функція Лафліна, хоча знайдена різниця між енергіями двох квазічастинок  Лафліна та складних ферміонів  зменшується зі збільшенням розміру системи.

	\keywords дробовий квантовий ефект Гола, сильно скорельовані системи, квазічастинкові збудження
	
\end{abstract}


\begin{thebibliography}{99}

\bibitem{R1} Collins~G.P., Phys. Today, 1997, \textbf{50}, 17, \doi{10.1063/1.882050}.

\bibitem{R2} Leinaas~J.M., Myrheim~J., J. Nuovo Cim. B, 1977, \textbf{37}, 1, \doi{10.1007/BF02727953}.

\bibitem{R3} Jain~J.K., Composite Fermions, Cambridge University Press, New York, 2007.

\bibitem{R4} Jain~J.K., Phys. Rev. B, 1990, \textbf{41}, 7653, \doi{10.1103/PhysRevB.41.7653}.

\bibitem{R5} Laughlin~R.B., Phys. Rev. Lett., 1983, \textbf{50}, 1395, \doi{10.1103/PhysRevLett.50.1395}.

\bibitem{R6} Jeon G.S., Jain J.K., Phys. Rev. B, 2003, \textbf{68}, 165346, \doi{10.1103/PhysRevB.68.165346}. 

\bibitem{R7} Bentalha~Z., Physica B, 2016, \textbf{492}, 27, \doi{10.1016/j.physb.2016.03.034}.

\bibitem{R8} Arovas~D.P., Schrieffer~J.R., Wilczek~F., Phys. Rev. Lett., 1984, \textbf{53}, 722, \doi{10.1103/PhysRevLett.53.722}.

\bibitem{R9} Jeon G.S., Graham~K.L., Jain~J.K., Phys. Rev. Lett., 2003, \textbf{91}, 036801, \doi{10.1103/PhysRevLett.91.036801}.

\bibitem{R10} Nayak~C., Simon~S.H., Stern A., Freedman~M., Das Sarma S., Rev. Mod. Phys., 2008, \textbf{80}, 1083, \\ \doi{10.1103/RevModPhys.80.1083}.

\bibitem{R11} Stern~A., Nature, 2010, \textbf{464}, 187, \doi{10.1038/nature08915}.

\bibitem{R12} Freedman~M.H., Kitaev~A., Larsen~M.J., Wang Z., Bull. Am. Math. Soc., 2003, \textbf{40}, 31, \\ \doi{10.1090/S0273-0979-02-00964-3}.

\bibitem{R13} Pachos Jiannis~K., Introduction to Topological Quantum Computation, Cambridge University Press, Cambridge, 2012.

\bibitem{R14} Ciftja~O., Physica B, 2009, \textbf{404}, 227, \doi{10.1016/j.physb.2008.10.036}.

\bibitem{R15} Ciftja~O., Phys. Lett. A, 2010, \textbf{374}, 981, \doi{10.1016/j.physleta.2009.12.017}.

\bibitem{R16} Morf~R., Halperin~B.I., Phys. Rev. B, 1986, \textbf{33}, 2221, \doi{10.1103/PhysRevB.33.2221}.

\bibitem{R17} Bentalha~Z., Moumen~L., Ouahrani~T., Cent. Eur. J. Phys., 2014, \textbf{12}, 511, \doi{10.2478/s11534-014-0476-5}.

\bibitem{R18} Ammar~M.A., Bentalha~Z., Bekhechi~S., Condens. Matter Phys., 2016, \textbf{19}, 33702:1--9, \\ \doi{10.5488/CMP.19.33702}.
\end{thebibliography}
\end{document}